\documentclass[fleqn,usenatbib]{mnras}
\usepackage[T1]{fontenc}
\usepackage{graphicx}
\usepackage{amsmath}
\usepackage{amssymb}
\usepackage{rotating}

\title{Inverted EEMD: a robust method to identify narrow absorption features form spectral data and cubes}

\author[Z.-Z. He et al.]{
Zhen-Zhen He$^{1,2}$, 
Guang-Xing Li$^{3}$\thanks{Contact e-mail: \href{mailto:gxli@ynu.edu.cn}{gxli@ynu.edu.cn}, \href{mailto:ligx.ngc7293@gmail.com}{ligx.ngc7293@gmail.com}}
\\
$^{1}$Research Center for Astronomical Computing, Zhejiang Laboratory, Hangzhou 311100, China\\
$^{2}$Department of Investigation, Henan Police College, Zhengzhou 450046, China\\
$^{3}$Department of Astronomy, Yunnan University, Kunming 650091, China\\
$^{4}$South-Western Institute For Astronomy Research, Yunnan University, Kunming 650091, China
}

\date{Accepted XXX. Received YYY; in original form ZZZ}
\pubyear{2022}

\begin{document}
\label{firstpage}
\pagerange{\pageref{firstpage}--\pageref{lastpage}}
\maketitle

\begin{abstract}
  Extracting information from complex data is a challenge shared by multiple frontiers of modern astrophysical research. Among those, analyzing spectra cubes, where the emission is mapped in the position-position-velocity space is a difficult task given the vast amount of information contained within. The cubes often contain a superposition of emissions and absorptions, where extracting absorption signatures is often necessary. One example is the extraction of
  narrow absorption structures in HI 21 cm emission spectra. These HI self-absorption (HISA) clouds trace the cold HI gas in interstellar space. 
  We introduce an automatic and robust method called the \emph{inverted EEMD} method to extract narrow features from spectral cubes. Our method is based on the EEMD method, an established method to decompose 1d signals. The method is robust and parameter-free, making it useful in analyzing spectral cubes containing localized absorption signals of different types. The inverted-EEMD method is suitable for the analysis of spectral cubes where it can produce a cube containing the absorption signal and one containing the unabsorbed signal, where cold clouds can be identified as coherent regions in the absorption map.  A Python implementation of the method is available at \url{https://github.com/zhenzhen-research/inverted_eemd_map}. 
\end{abstract}

\begin{keywords}

Algorithms, ISM: clouds, ISM: atoms, method: data analysis
\end{keywords}

\section{Introduction}

Most of our understanding of the universe comes from analyzing emissions from
astronomical objects. Among the data, one type that is hard to
analyze is the spectral cubes. In these cubes, the objects, such as gas clouds or galaxies,
are mapped on the sky, with spectral information available at every
location. The data is often provided in a cube-like data structure in the position-position-velocity spaces, called 
spectral cubes. 
The vast amount of information contained in these cubes demands automatic methods,
yet the complex nature of the spectral lines profiles, where emission and
absorption signatures often appear superimposed, making the developments of
these automatic methods are challenging.

One type of such spectral feature is the narrow,
localized absorption signatures, caused by the presence of cold gas along the
line of sights.
These features are not easy to separate. However, when the background is smooth and the absorption is localized, it is possible to
separate these absorption features. 
One well-known example of a type of absorption signature is the HI
Narrow Self-Absorption (HINSA) lines \citep{1954AJ.....59..324H,1955ApJ...121..569H,2003ApJ...585..823L} produced by the cold HI clouds against warm HI gas.
The warm gas emits radiation at 21 cm, while the cold cloud absorbs radiation at this wavelength.
Understanding the relationship between different gas phases is necessary for understanding gas circulation in the Galactic ecosystem.

With an increasing amount of data being produced by recent surveys,  developing effective algorithms to separate narrow absorption features from observational data cubes has become an urgent task. In the past, methods have been developed where the spectrum is assumed to be a superposition of multiple positive and negative Gaussian components, then the cold gas might be traced by the negative Gaussian \citep{2000ApJ...540..851G,2005ApJ...626..195G,2005ApJ...626..214G}. The map presented by the authors appears to have a lack of dynamic range.  Others \citep{2005ApJ...626..887K} have attempted to extract absorption using the 2nd-order duratives along the frequency axis. However, this approach requires data with a high signal-to-noise ratio. {\bf Another common problem with the previous works is the implementation of the algorithms is often not available. This makes it difficult to judge the effectiveness of the methods, which also hinders widespread adoption.}

We propose an Ensemble
Empirical Mode Decomposition (EEMD)-based method \citep{1998RSPSA.454..903H, article} to separate narrow absorption features from spectral cubes. The EEMD method
is an empirical method developed to decompose signals into components of
different frequencies. Compared to other approaches such as the
wavelet transform, the EEMD approach is robust against
non-linear features in data. This makes the method particularly useful for
HI emission profiles, which are complex and can be fast-varying. Based on the
EEMD, we propose an automatic and robust method to extract narrow absorption features from HI
emission maps. The method performs decomposition to the individual spectra. When analyzing data cubes, we apply the method to each spectrum, one can obtain a cube containing the absorption signal and another one containing the unabsorbed signal. If the materials that introduce absorption have coherent distribution, they will show up as coherent regions in the absorption cubes. The absorbing cubes are an ideal starting point for studying these cold clouds. 
In this paper, we present the method and demonstrate its performance by applying it to data from the THOR survey, which contains 21cm radio observations towards the Milky Way disk, to extract absorption signatures produced by the HISA.

\section{Method} \label{sec:method}
\subsection{Procedure overview}
In observations, the HI emission data is presented as 3D cubes in the position-position-velocity (PPV) space.
At each sky position, there is a spectrum I$(v)$ with $v$ being the velocity.
The method is designed to extract 3D coherent absorption structures in a data cube traced by narrow absorption features in the 1D spectrum.
An overall finding process of our method is summarized in Fig. \ref{fig:dec_proc}.

We start by decomposing emission spectrum I$(v)$ using EEMD algorithm, then extracting absorption features by inverse decomposed components. The whole process is thus called the Inverted EEMD method.

By traversing all spectra of the data cube using the Inverted EEMD method, we reconstruct the 3D absorption feature cube and background cube.
The coherent 3D structures in the absorption cube might be identified as possible HISA clouds.

A detailed explanation of the process is as follows.

\subsection{Inverted EEMD}\label{sec:emd}
The Ensemble Empirical Mode Decomposition (EEMD) method is a data-adaptive technique to decompose 1D signal into several physically meaningful components -- the so-called intrinsic mode functions (IMFs) -- and residual signal \citep{1998RSPSA.454..903H,article}.
The sum of all IMFs and the residual signal would be the original signal.

\begin{figure*}
\centering
\includegraphics[width=1.2\textwidth,angle=90]{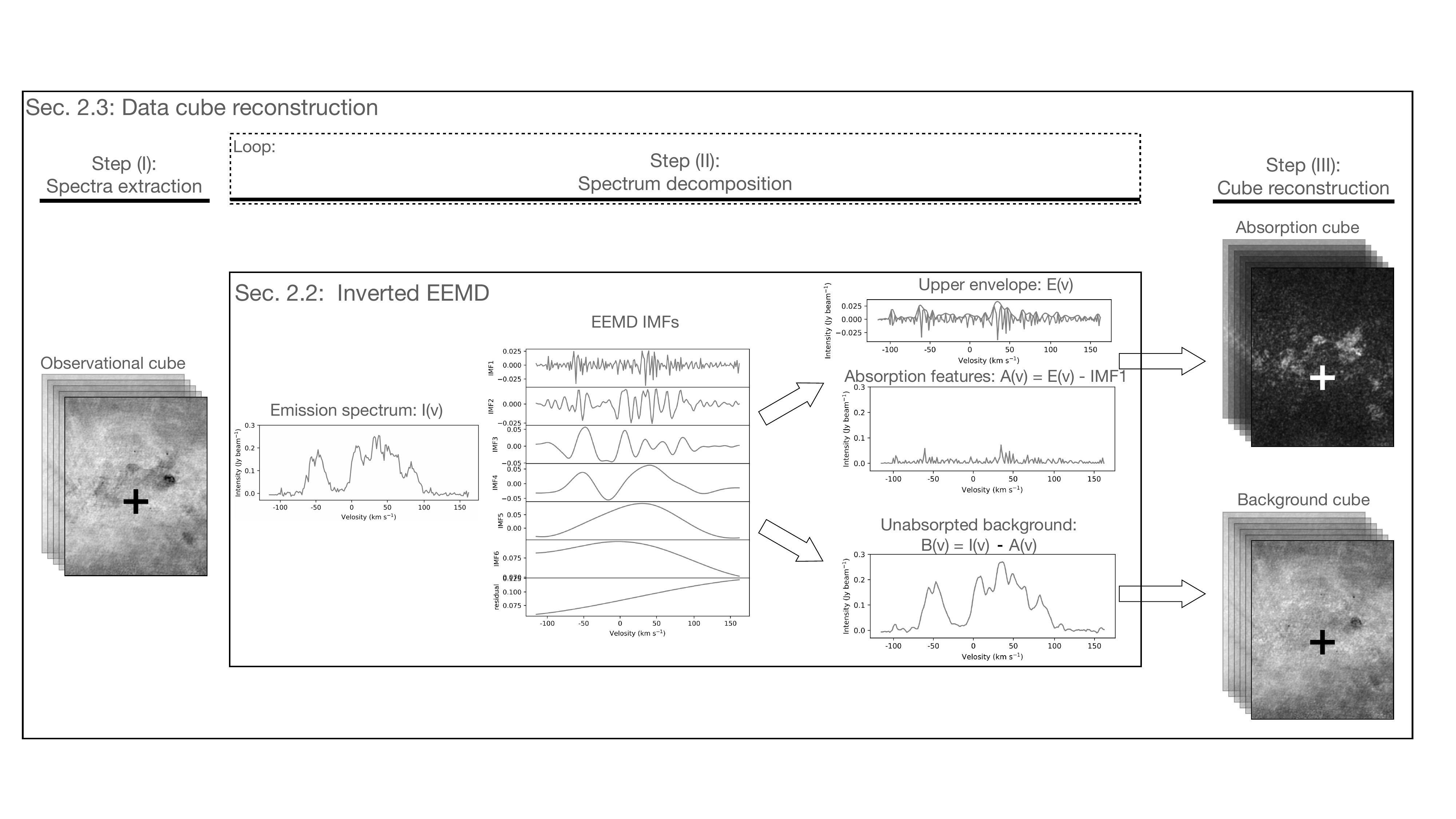}
\caption{A flowchart of the inverted EEMD method in extracting absorption features from an emission map. The spectrum decomposition is introduced in Sec. \ref{sec:emd}. The data cube decomposition procedure can be divided into three steps: Spectra extraction, Spectrum decomposition, and Cube reconstruction. A detailed introduction of the procedure is given in Sec. \ref{sec:procedure}.}
\label{fig:dec_proc}
\end{figure*}

As shown in Fig. \ref{fig:dec_proc}, for a emission spectrum I$(v)$ extracted from
HI emission map, the inverted EEMD process of extraction absorption features is as follows:
\begin{enumerate}
  \item Apply the EEMD algorithm to decompose I$(v)$ into IMFs. The IMF1 contains the fastest-varying component of the signal, the narrow features we are interested in may be hidden within it.
  \item Find local maxima values of the IMF1.\label{i2}
  \item Create the upper envelope E$(v)$ from an array of those maxima values by cubic interpolation.
  \item Calculate absorption features by inverse IMF1: A$(v)$ = E$(v)$ - IMF1 (According to our definition, A$(v) > 0$ is related to an absorption signal), then the unabsorbed signal (background) can be obtained as B$(v)$ = I$(v)$ - A$(v)$. \label{i5}
 \end{enumerate}
Because the absorption features are obtained by inverting the IMF1 obtained through the EEMD, the method is named \emph{Inverted EEMD}.

Toward each input spectrum I$(v)$, we obtain absorption features A$(v)$ and background signal B$(v)$. The possible absorption features are contained in A$(v)$. One can also apply the steps \ref{i2} - \ref{i5} to  IMFs of higher orders, to extract absorptions of larger widths. This functionality is implemented in the version of the code available online.

\subsection{Data cube reconstruction}\label{sec:procedure}
The inverted-EEMD method is a method to decompose 1d spectra into absorption spectra and an unabsorbed signal. The major advantage of the method is its robustness, and the best use case is to apply it to data cubes in a pixel-by-pixel way, producing a cube containing all the absorption signatures, and one containing the unabsorbed signals. 

The data cube decomposition procedure can be divided into three steps: Spectra extraction, Spectrum decomposition, and Cube reconstruction.
The details are:
\begin{enumerate}
\item Spectra extraction: The observational HI 21 cm emission data in ($l,b,v$) space is placed in a 3D cube with a size of N$\times$M$\times$L. Here $l$ and $b$ are galactic latitude and longitude, respectively.
Each pixel of the $l-b$ plane corresponds to a position on the sky plane. There are N$\times$M pixels, so there are N$\times$M spectra I$(l_{i}, b_{j}, v)$ of length L.
\item Spectrum decomposition: Toward each location, we decompose spectrum I$(l_{i}, b_{j}, v)$ using the Inverted EEMD method to obtain absorption features A$(l_{i}, b_{j}, v)$ and background spectrum B$(l_{i}, b_{j}, v)$, see Sec.\ref{sec:emd}.
\item Cube reconstruction: Iterate through step (ii), then place the results A$(l_{i}, b_{j}, v)$ and B$(l_{i}, b_{j}, v)$ into data cubes to create absorption feature cube A$(l,b,v)$ and background cube B$(l,b,v)$, respectively.
\end{enumerate}

In this step, we put in a 3D data cube and obtain absorption features cube A$(l,b,v)$ and background cube B$(l,b,v)$. The coherent 3D structures in A$(l,b,v)$ might be identified as possible cold clouds.

\begin{figure}
  \centering
  \includegraphics[width=0.5\textwidth]{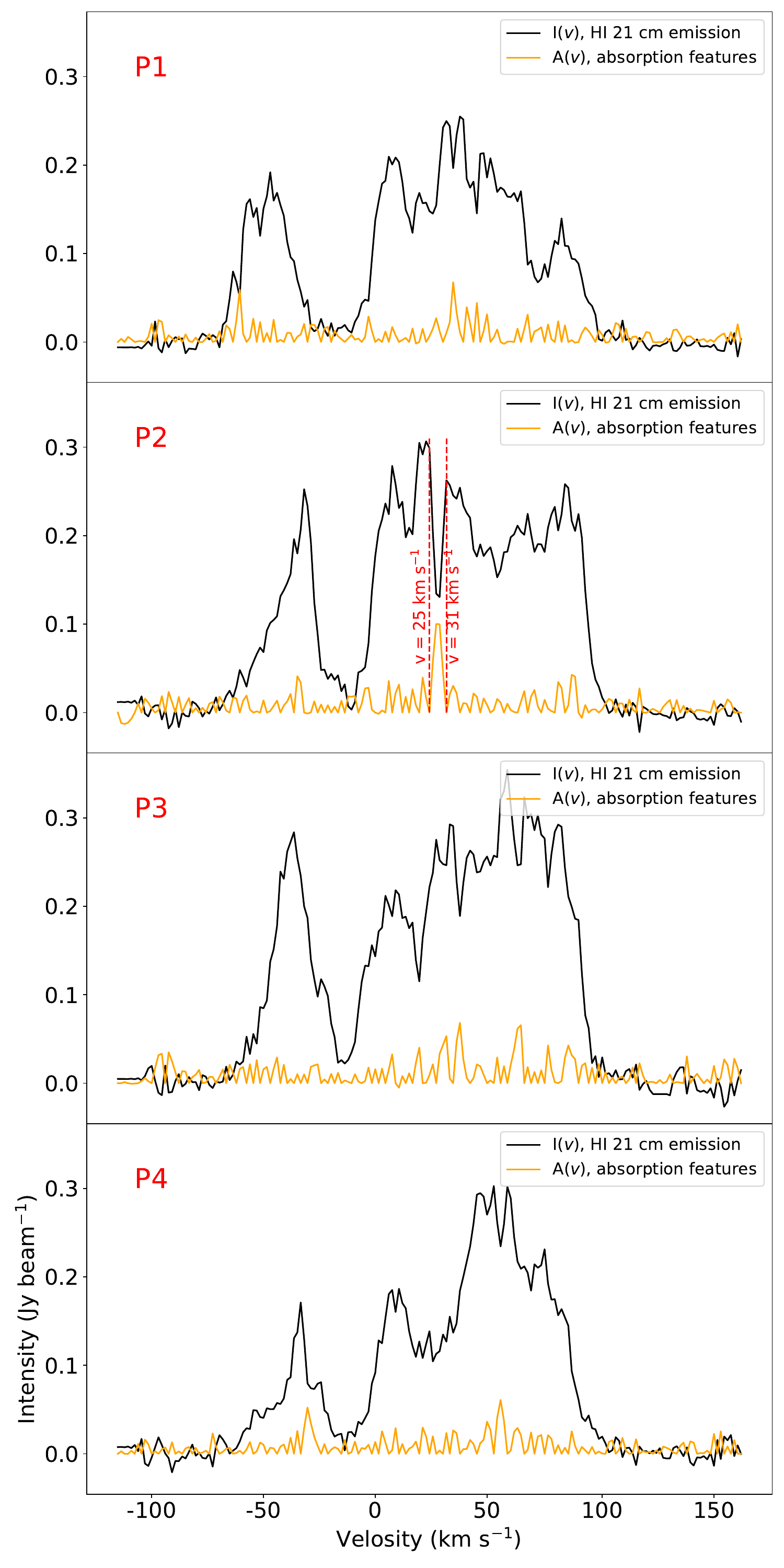}
  \caption{The HI emission spectra (black) and absorption features (orange) were derived using the inverted EEMD method.
  The four emission spectra are extracted at four different positions of the THOR survey (see Fig \ref{fig:fig}).
  A possible HISA feature in the P2 spectrum with a velocity range of $\sim$ 25 - 31 km s$^{-1}$ was labeled.}\label{fig:spec}
  \end{figure}

\begin{figure}
\centering
\includegraphics[width=0.45\textwidth]{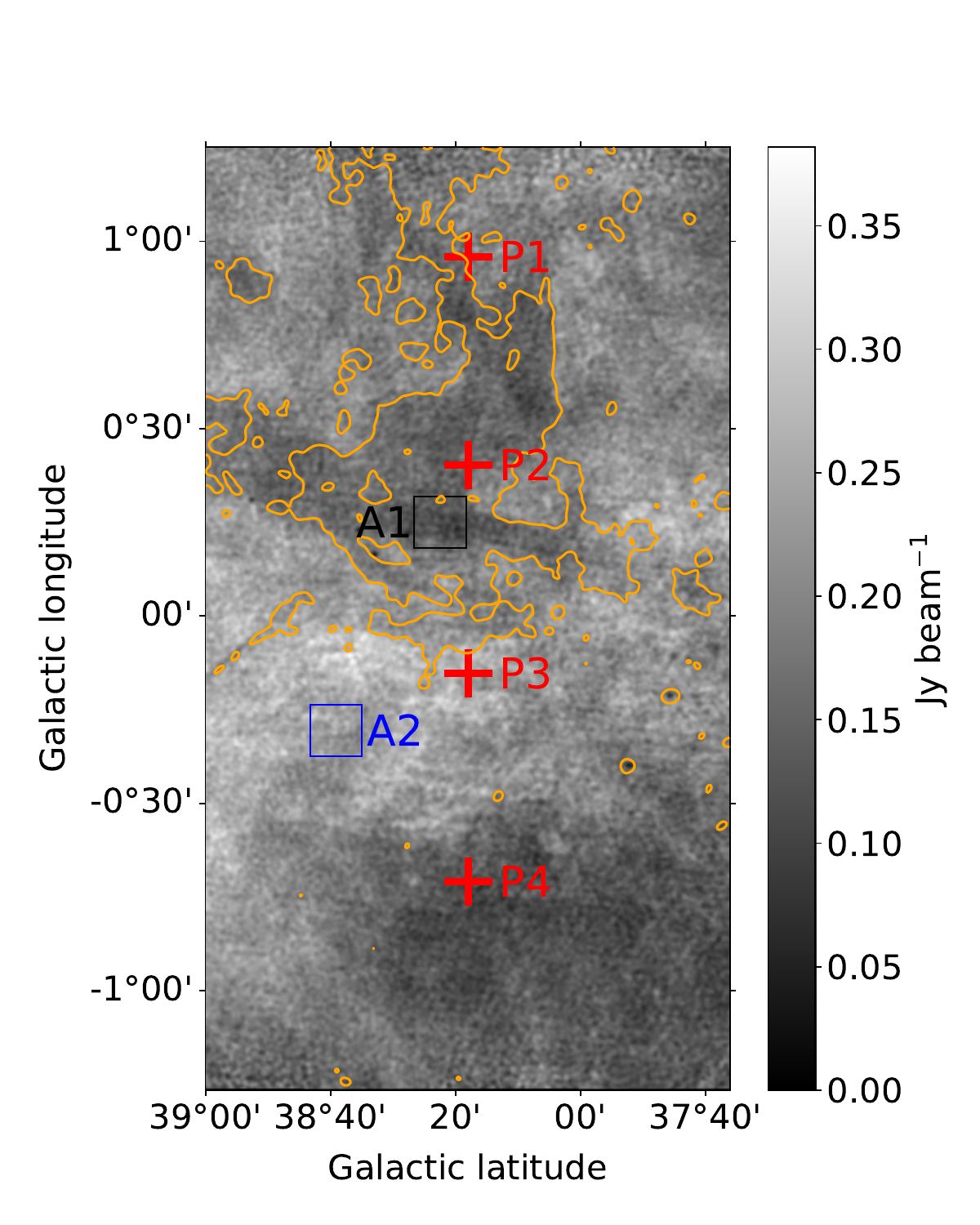}
\caption{A possible HISA cloud was identified by the inverted EEMD method. The background is an integrated map of HI 21 cm emission between a velocity range of 25-31 km s$^{-1}$ from the THOR survey. The boundary of the possible HISA cloud is outlined using the orange contours. The red plus symbols mark four different positions where we extract emission spectra.
The black and blue squares are regions where we take the box-averaged spectra.}\label{fig:fig}
\end{figure}

\begin{figure*}
\centering
\includegraphics[width=\textwidth]{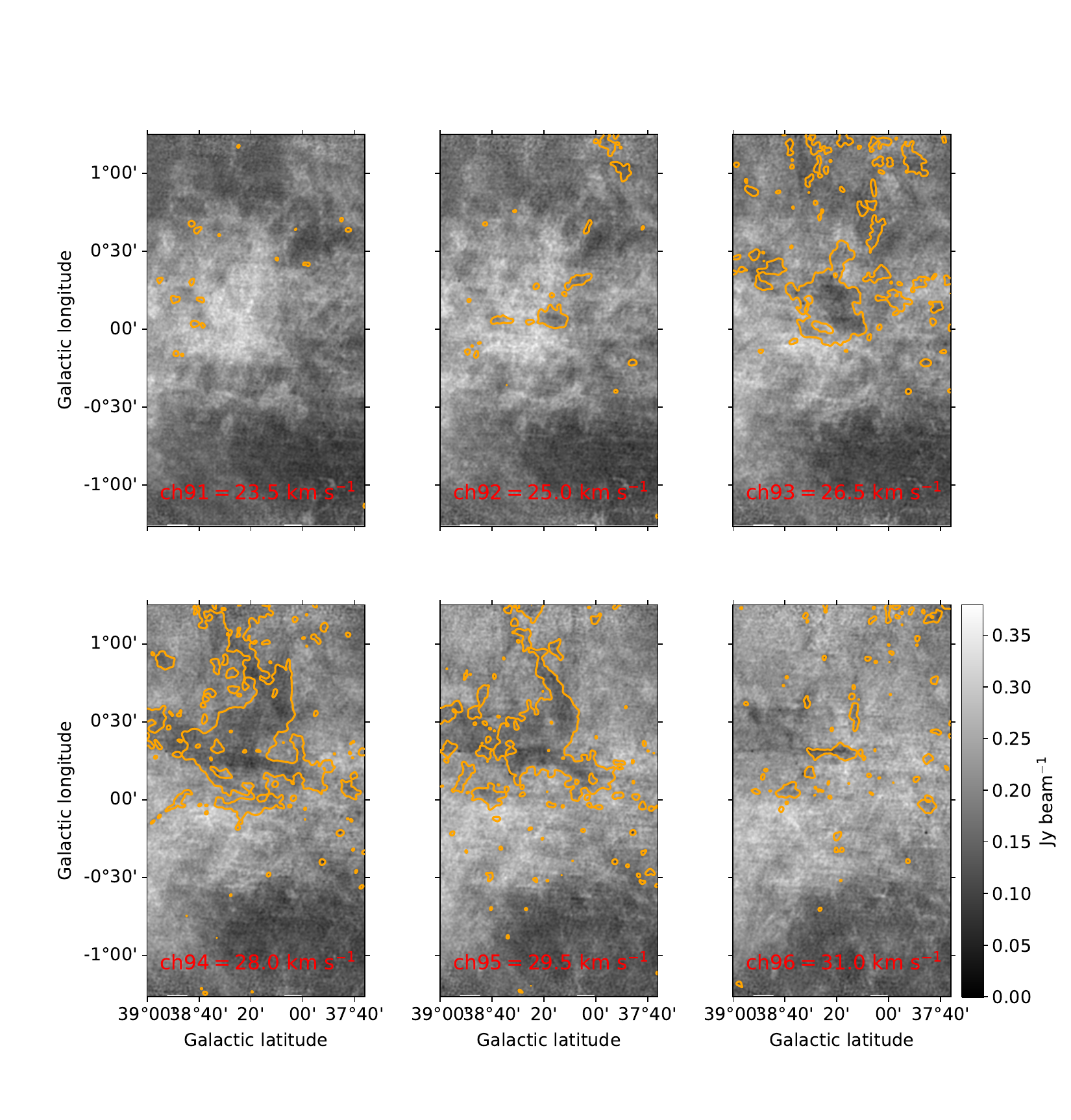}
\caption{The channel map of a HISA cloud. The background is HI emission data. Contours outline the HISA cloud with intensity $\geq$ 0.03 Jy beam$^{-1}$ in the absorption cube A$(l, b, v)$.}\label{fig:ch}
\end{figure*}

\begin{figure*}
\centering
\includegraphics[width=\textwidth]{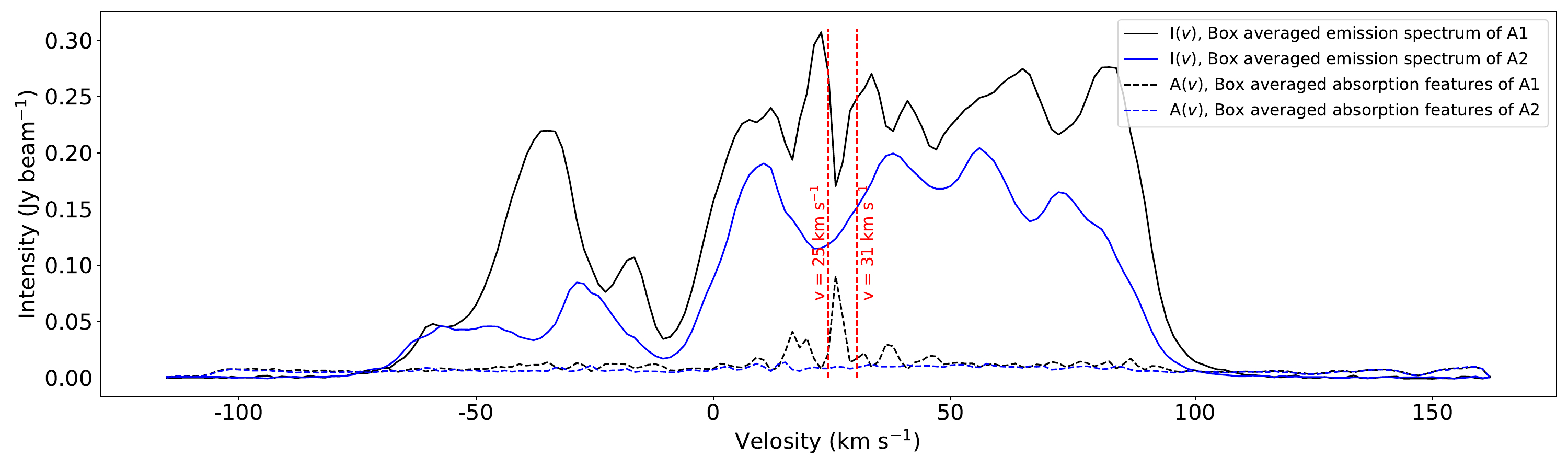}
\caption{Box-averaged emission and absorption features of area A1 and A2 are indicated in Fig. \ref{fig:fig}. Area A1 shows an obvious HISA-like feature in the spectrum between the velocity range of 25-31 km s$^{-1}$, which is marked using red dashed lines, but not in area A2.}\label{fig:box_spec}
\end{figure*}

\section{Application} \label{sec:application}

\subsection{Data} \label{sec:data}
We use the HI/OH/Recombination (THOR) line survey of the Milky Way \footnote{We download the HI data from website: \url{https://www2.mpia-hd.mpg.de/thor/Overview.html}} as our test dataset \citep{2016A&A...595A..32B,2020A&A...634A..83W}.
The observations are conducted using the Very Large Array (VLA), to demonstrate the capability of our method in extracting HISA clouds. 
The THOR data have an angular resolution of about 25$\arcsec$ and a velocity range of about -100 - 150 km s$^{-1}$ with a resolution of about 1.5 km s$^{-1}$.
A subdataset of the THOR survey covering l = $\sim$37.5$^{\circ}$ - 39$^{\circ}$ and |b| $\leq$ 1.25$^{\circ}$ has been selected as our test field.
In Fig. \ref{fig:fig}, the selected fields are shown, where the background is the integrated intensity map of observational data.

\subsection{Results}
\subsubsection{Spectrum}\label{sec:spectrum}
We test the capability of the inverted EEMD in detecting HISA-like features in the real observational spectra.

Four emission spectra are extracted as test signals from the THOR data cube, they are indicated in Fig. \ref{fig:fig} by red plus symbols.
The emission spectra I($v$) as well as the absorption features A($v$) are plotted in Fig. \ref{fig:spec}.

From Fig. \ref{fig:spec}, most of the narrow features upon the emission spectra I($v$) are indicated by the absorption features A($v$), and the intensity of A(v) is proportional to the ``absorption intensity''.
The high intensity of A(v) indicates possible HISAs in the emission spectra.
In the absorption features of the P2 position, we have labeled a feature with a max intensity of about 0.1 Jy beam$^{-1}$ between the velocity range of 25-31 km s$^{-1}$, this feature is believed to be a possible HISA.

\subsubsection{Data cube}\label{sec:cube}

The coherent 3D structures in the absorption cube A$(l, b, v)$ are labeled as possible HISA clouds.
In this section, we aim to show the coherent structures having HISA-like features in the observational spectrum.

To identify the coherent structures in A$(l, b, v)$, the image of each velocity channel is smoothed.
In addition, to extract obvious HISA sources, we filter noise from A$(l, b, v)$.
A threshold of 5 times the noise level ($\sim$ 0.03 Jy beam$^{-1}$) has been chosen to focus on the most significant structures. 

A 3D cloud structures was labeled in A$(l, b, v)$, the possible HISA cloud centered around l = 30$^{\circ}$20$'$, b = 0$^{\circ}$30$'$, and v = 28 km s$^{-1}$ with a velocity range between 25-31 km s$^{-1}$ (see the spectral from Fig. \ref{fig:spec} where this HISA feature is indicated.).
In Fig. \ref{fig:ch}, the cloud slice outlined by contours is shown as a channel map, the cloud is overlapped on HI 21 cm emission.

We projected the 3D cloud structures into 2D for viewing in Fig. \ref{fig:fig} and the cloud is outlined by orange contours.
Within the outlined region, we have chosen a small area A1 to show the box-averaged emission spectrum in Fig. \ref{fig:box_spec}, as well as an area A2 outside the region.
A HISA feature is shown in A1 spectrum between velocity range of $\sim$ 25 - 31 km s$^{-1}$.
Compared to this, the box-averaged spectrum of A2 does not show an obvious absorption feature. This proves our goal: coherent structures in A$(l, b, v)$ have HISA-like features in the spectrum I($v$).

\section{HISA on a Giant Molecular Filament}
In this section, we examine the HISA detection in detail within a giant molecular filament GMF38a, which is distributed in the velocity range of 50-60 km s$^{-1}$ \citep{2014A&A...568A..73R}. 

The strong continuum sources can also result in absorption features in the HI emission data, which are confused with HISA absorptions.
The continuum map at 1060 MHz that was observed simultaneously with HI line emission in the THOR survey is used to identify those regions \citep{2016A&A...595A..32B,2020A&A...634A..83W}.
We regard the regions where the continuum flux is greater than 3$\sigma$ of the continuum map as continuum source regions and set those regions of absorption cube A($l$, $b$, $v$) to 0 to ensure robustness.

\begin{figure*}
\centering
\includegraphics[width=1.3\textwidth,angle=90]{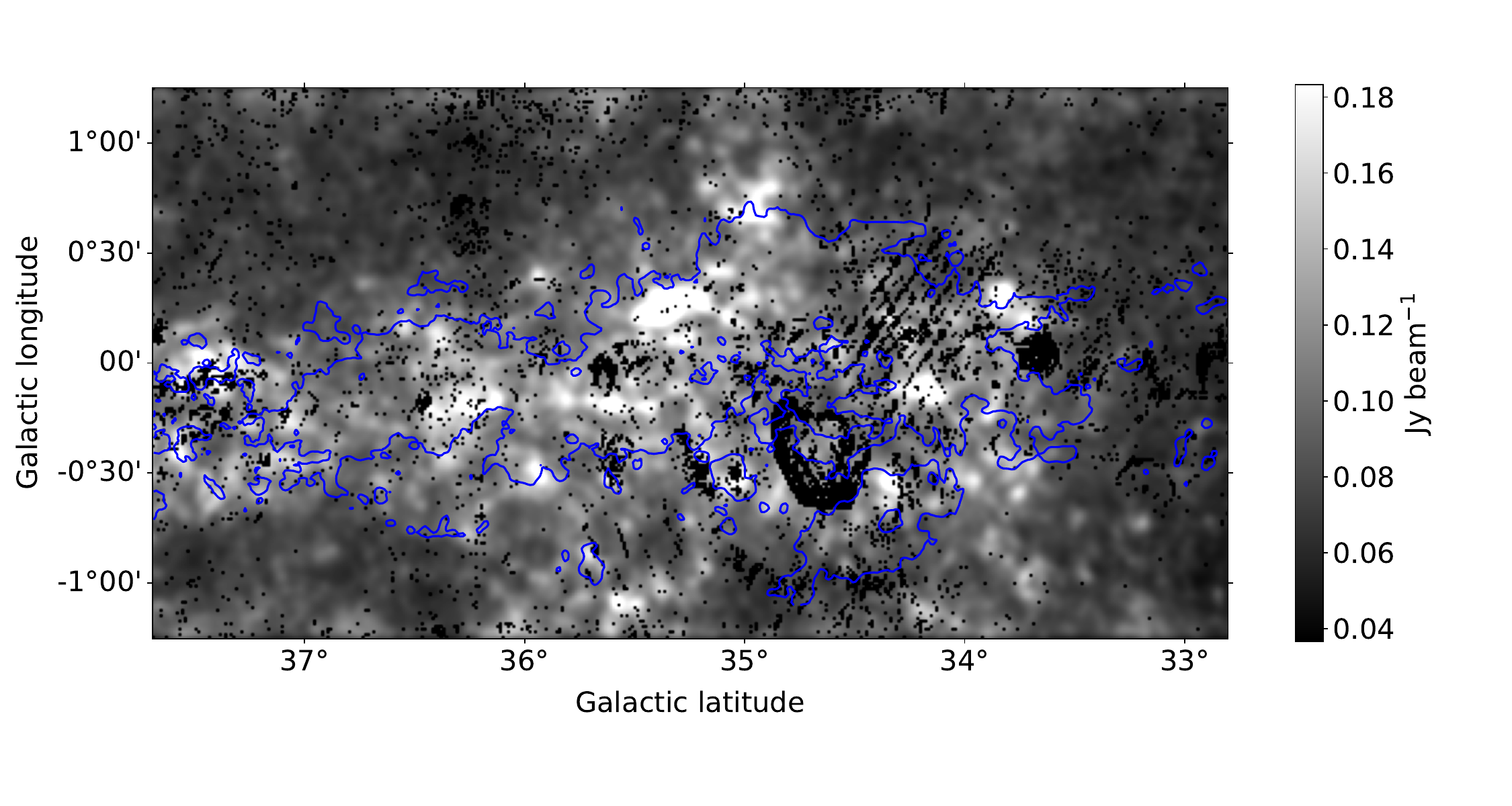}
\caption{An integrated map of absorption features cube A$(l, b, v)$ obtained by the inverted EEMD method between velocity range of 50 -60 km s$^{-1}$.
The blue contours are the GMF38a $^{13}$CO integrated emission at levels of 5 K km s$^{-1}$. The $^{13}$CO data comes from GRS survey \citep{2006ApJS..163..145J}.}\label{fig:cmp2wang}
\end{figure*}

The Fig. \ref{fig:cmp2wang} shows an integrated map of the absorption cube obtained by our method and GMF38a $^{13}$CO emission from the GRS survey as contours in the velocity range of 50 - 60 km s$^{-1}$ \citep{2006ApJS..163..145J}.
Our HISA map appears to trace the $^{13}$CO filament better compared to \citet{2020A&A...634A.139W} where HISA is detected only in a few isolated stripes.
Through visual inspections, we have verified the existence of absorptions towards most of the filament body. Our detection of HISA feature toward a larger fraction of the filament body is consistent with astro-chemical models of cloud formation, where high-temperature HI gas gradually moves from warm to cold phase through cooling mechanisms \citep{2003ApJ...587..278W}.
As the temperature decreases, molecules (e.g., H$_2$, CO) gradually form within atomic clouds (e.g., H, C, O) \citep{2018ApJ...867...13Z}. Thus, it is natural to expect a CO cloud to contain a fraction of HI gas tradable using HISA.

Compared to the results in  \citet{2020A&A...634A.139W}, the performance of our method is much improved as it can detect HISA towards a much larger area. This advantage arises from the model-free nature of the EMD method. The package we offer 
 does not require inputs such as the centroid velocity, and is more robust and efficient towards large datasets. 

\section{Conclusion} \label{sec:Conclusion}
We propose the inverted Ensemble Empirical Mode Decomposition (inverted
EEMD) method to identify possible HISA features automatically from HI
emission spectra. The method is based on the decomposition of the initial signal using the EEMD (Ensemble Empirical Mode Decomposition) and the extraction of the
absorption signatures as the differences between a signal and its upper envelope. The method is a model-free, noise-resistant method and is particularly useful in extracting narrow absorption features in spectral data of large volumes.

By applying our method to survey data cubes, we provide
cubes that contain the narrow absorption features as well as cubes containing the unabsorbed background. Coherent 3D structures
can be identified from the absorption cubes, which should correspond to cold clouds. The method is
particularly suitable to be applied to large, homogeneous datasets
toward galaxy ISM to perform systematic studies. 
The method is available at \url{https://github.com/zhenzhen-research/inverted_eemd_map}.

\section*{Acknowledgements}
GXL acknowledges support from NSFC grant No. 12273032 and 12033005.


\bibliographystyle{mnras}
\bibliography{ref}
\end{document}